# Adjoint optimization of polarization-splitting grating couplers

**PENG SUN,**[1, †,*] **THOMAS VAN VAERENBERGH,**[2] **SEAN HOOTEN,**[1] **AND RAYMOND BEAUSOLEIL**[1]

[1]*Hewlett Packard Labs, 820 N McCarthy Blvd, Milpitas, CA 95035, USA*
[2]*Hewlett Packard Labs, HPE Belgium, B-1831 Diegem, Belgium*
[†]*Currently at NVIDIA Corporation, 2788 San Tomas Expy, Santa Clara, CA 95051*
*\*psun@outlook.com*

**Abstract:** We have designed a polarization-splitting grating coupler (PSGC) in silicon-on-insulator (SOI) that has 1.2 dB peak loss in numerical simulations, which is the best simulated performance of PSGCs without a bottom reflector to the best of our knowledge. Adjoint method-based shape optimization enables us to explore complex geometries that are intractable with conventional design approaches. Physics-based process-independent knowledge of PSGCs is extracted from the adjoint optimization and can be transferred to other platforms with a minimum of effort.

## 1. Introduction

Polarization-splitting grating couplers (PSGCs) interface silicon photonic integrated circuits with single-mode optical fibers that have random polarization states [1-4]. A PSGC consists of a two-dimensional (2D) array of scattering elements that scatter light from an optical fiber into two orthogonal waveguides, or vice versa. The scattering elements, which typically have the quantity of ~800 to match the mode size of an SMF-28 fiber, are arranged on collinear lattices or elliptical lattices to generate planar phase front in the scattered mode to match that of the optical fiber. For collinear lattices, the optical fiber mode is converted to planar phase front in the silicon slab, and then a pair of long waveguide tapers convert the slab mode (~10 um in width) to a pair of orthogonal single-mode silicon rib waveguides (~400 nm in width). For elliptical lattices, the optical fiber mode is converted to cylindrical phase front in the silicon slab that are focused at the two focal points of the elliptical lattices, where a pair of single-mode silicon rib waveguides are placed.

Optimizations of PSGCs should include two objectives: to maximize coupling efficiency and to minimize polarization dependent loss. First, the coupling efficiency can be improved by tailoring the grating's scattering strength to match the fiber mode profile, which is commonly referred to as apodization. The ideal scattering strength of one-dimensional (1D) gratings that generate gaussian mode has closed-form solution as shown in eqn. (4) in [5], which requires weak scattering at the beginning of the grating, and strong scattering at the end. The layout of a PSGC possesses mirror symmetry to split the incident light from the optical fiber into two identical, orthogonal silicon rib waveguides, and therefore the apodization of the grating must be described by a mirror-symmetric 2D profile. For PSGCs, intuitively the scattering strength should have similar characteristics that the scattering is weaker at the edge of the grating area and stronger at the center. The scattering strength can be modulated by the scatterers' aspect ratio [6], that is, to stretch a polygon along one silicon waveguide and compress it along the orthogonal waveguide while preserving the area of the polygon such that the phase matching condition is not perturbed. Higher aspect ratio generates weaker scattering for light propagation along the stretching axis, and stronger scattering along the compressing axis. Second, the optical fibers are often tilted/polished at a small angle to suppress reflection of the gratings and reflection at the fiber-chip interface. As a consequence, breaking of the circular symmetry in the fiber-grating mode overlap introduces disparity in the coupling efficiencies between the two orthogonal polarization states. It has been shown that the polarization-dependent loss (PDL) and polarization-dependent wavelength (PDW) can be effectively modulated by stretching the scattering elements along the S- and P-polarization axes [7-9].

A thorough optimization of PSGCs will require designers to choose the aspect ratio and the shape parameters for each and every scatterer, which is infeasible given the sheer number of scatterers. The adjoint method enables exploration of high-dimensional parameter spaces by calculating the gradient of a Figure-of-Merit (FOM) function of a linear system with fixed and low computational cost, regardless of the number of parameters [10-13]. The highly complicated PSGC problem is uniquely suited for adjoint method. In this work, we designed PSGCs with adjoint method, and demonstrated peak coupling loss of 1.2 dB without a bottom reflector. Section 2 details the parameterization, optimization, and validation of the PSGCs. We also present our efforts to interpret and understand the physical insights behind the adjoint-optimized PSGCs. Section 3 studies the dimensionality reduction and manufacturability of the adjoint-optimized designs. Section 4 concludes the paper with discussions on future works.

## 2. Adjoint optimization of PSGCs

We study PSGCs based on a collinear lattice, and the technique developed in this work can be generalized to PSGCs on an elliptical lattice. An example PSGC on a collinear lattice is shown in Fig. 1(a). The PSGC is excited from a waveguide source on one of the two symmetric arms. Scattered fields are captured by a monitor on top of the grating area and matched to a tilted gaussian mode to compute the coupling efficiency. The coupling efficiency of a fiber-excited PSGC is identical to that of a waveguide-excited PSGC of the same design, as dictated by the reciprocity of the device. The tilted gaussian mode, which emulates the mode of an angle-polished optical fiber, is centered at the intersection of the two orthogonal waveguide arms. Projection of the fiber wavevector on the grating, as well as the S- and P-polarization vectors, are schematically shown in Fig. 1(a). Figure 1(b) shows the example of a circular scatterer under the transformation of aspect ratio AR=1.2, where all the points are scaled by a factor of AR along X axis and 1/AR along Y axis.

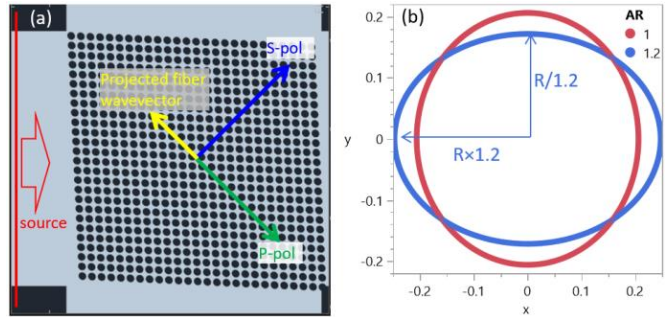

Fig.1 (a) Top view of a PSGC on a collinear lattice. (b) Illustration of transforming a circle by aspect ratio.

We expect scatterers at different locations to have different aspect ratio. To optimize apodization of the PSGC, we synthesize the 2D map of aspect ratio across the grating area using generalized Fourier series. We choose 2D Chebyshev polynomials as the basis functions for two reasons. First, as an intuitive guess inspired by the ideal 1D scattering strength, the scattering strength of 2D gratings should increase monotonically from the starting edge to the center, and therefore can be approximated by polynomials. Second, Chebyshev polynomials yield the best polynomial approximation of all the polynomials with identical principal coefficients [14]. Figure 2 shows the contour maps of 2D Chebyshev polynomials on a unit square $[-1,1]\times[-1,1]$ up to the 2$^{nd}$ order, with mirror symmetry enforced on the diagonal axis. The first two terms with p=0, q=0, and p=1, q=0 determine a linearly ramped AR profile along the two waveguides, and intuitively they should play a more important role than other coefficients.

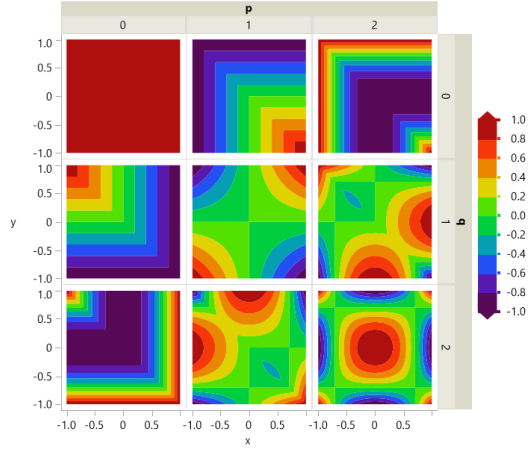

Fig.2 2D Chebyshev polynomials up to the $2^{nd}$ order, with mirror symmetry enforced on the diagonal axis.

For shape optimization of the scatterers, we synthesize each individual polygon with sine and cosine basis functions in polar coordinates as shown in eqn. (1), where $c_n$ and $s_n$ are the trigonometric coefficients and $r_0$ is the DC component:

$$r(\theta) = r_0 + \sum_{n=1}^{N}(c_n \cos(n\theta + \pi) + s_n \sin(n\theta + \pi)) \qquad (1)$$

Area of the polygon is given by eqn. (2):

$$A = \frac{1}{2}\int_0^{2\pi} r^2(\theta)d\theta = \pi r_0^2 + \frac{\pi}{2}\sum_{n=1}^{N}(c_n^2 + s_n^2) \qquad (2)$$

Fill Factor of the grating, which is defined as the ratio of the polygon area over the unit cell area, is forced to be constant regardless of the polygon shape to generate uniform phase front. The DC component $r_0$ needs to be calculated for each polygon shape, as shown in eqn. (3):

$$r_0 = \sqrt{\frac{pitch^2 \times FF}{\pi} - \frac{1}{2}\sum_{n=1}^{N}(c_n^2 + s_n^2)} \qquad (3)$$

The first 6 orders of sine and cosine functions in polar coordinates, with coefficient of 40 nm on top of a 200-nm-radius circle for all the basis functions, are illustrated in Figure 3. Each of the basis functions serves a specific purpose in modulating the light scattering. For example, the 2nd order cosine function stretches the scattering elements along one of the two orthogonal waveguides, and primarily increase or decrease the scattering intensity [6]. The $2^{nd}$ order sine and $4^{th}$ order cosine stretch the polygon towards the four corners of the unit cell along the S- and P-polarization in the optical fiber, which can modulate the grating's PDW/PDL [7]. We expect the $2^{nd}$ order sine and $4^{th}$ order cosine to play an important role in modulating the PDL/PDW. The $4^{th}$ order sine function stretches the polygon towards the four corners of the unit cell but off by an angle of 22.5°, and therefore we do not expect it to be as important as the $4^{th}$ order cosine function in modulating the PDL/PDW. We stop at the highest order of 6, as higher order trigonometric basis functions correspond to fine features and sharp angles in the scattering elements, which will have worse manufacturability and diminishing impact on the optical performance as the order increases.

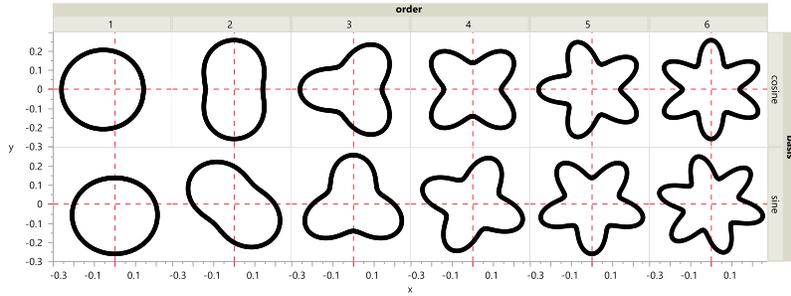

Fig. 3 Scatterer shapes of the first 6 orders of sine and cosine functions. The 2$^{nd}$ order sine and 4$^{th}$ order cosine should modulate PDW/PDL more effectively.

We expect polygons at different locations to have different shapes, therefore the 12 trigonometric coefficients will vary across the grating area. We synthesize the 2D maps for each of the trigonometric coefficients using 5$^{th}$ order 2D Chebyshev polynomials. For the 6 sine and 6 cosine functions, it takes (6+6)×25=300 parameters. Together with the 5×5=25 parameters for the aspect ratio, it takes 325 parameters in total to fully describe a PSGC of this design.

We use the open-source adjoint method toolkit EMopt [15], which has a built-in 3D FDTD solver. The FDTD solver of EMopt employs continuous-wave monochromatic sources, and therefore an optimization at multiple wavelengths requires one forward and one adjoint FDTD at each wavelength point. A broadband pulse source can be used to solve Maxwell's equations at multiple wavelength points in one FDTD, but the memory requirement for the Fourier transforms of EM fields in 3D design-perturbed regions, instead of 2D monitor planes, is excessive for large, complicated geometries like the PSGCs. In this study, the FOM of the PSGC optimization problem is defined as weighted sum of 3 single-wavelength FOMs at 1305 nm, 1310 nm, and 1315 nm respectively. Three wavelength points are used to construct the FOM in order to avoid degenerate solutions which have the S- and P-polarization coupling efficiency spectra crossing over at the one or two wavelength points where the FOM is defined. Each single-wavelength FOM is defined as the sum of two terms: the lower coupling efficiency of S- and P-polarization, minus the square of the difference of the S- and P-polarization coupling efficiencies. The first term uses log-sum-exp function to implement the soft minimum function [16] to preserve differentiability. The second term is a proxy of the PDL, and the square function is used in lieu of absolute function to preserve differentiability. To yield symmetric coupling spectra and to penalize PDL on the two sides more than at the center wavelength, we choose the wavelength weights as $c_{1305}:c_{1310}:c_{1315}=2:1:2$. To equalize the impact of loss and PDL at each wavelength point and to ensure that the optimized PDL is a small fraction of the loss, the PDL weight parameter is chosen to be $\zeta=1000$. The total FOM is shown in eqn.(4).

$$\begin{aligned}FOM = &\ c_{1305}[\min(S_{1305}, P_{1305}) - \zeta(S_{1305} - P_{1305})^2] \\ &+ c_{1310}[\min(S_{1310}, P_{1310}) - \zeta(S_{1310} - P_{1310})^2] \\ &+ c_{1315}[\min(S_{1315}, P_{1315}) - \zeta(S_{1315} - P_{1315})^2]\end{aligned} \quad (4)$$

We design the PSGCs for an SOI platform that has a 270 nm thick silicon layer on top of a 680 nm thick buried oxide layer. The scattering elements and the rib waveguides are formed by etching 140 nm into the top silicon layer. The PSGCs are designed for coupling to 8° angle polished single-mode fibers, which have MFD of 9.2 um at 1310 nm. In this study, we choose collinear lattice to save simulation time and to focus on the optimization of the grating area. The physics insights gained from the adjoint optimization can be applied to design PSGCs on elliptical latices.

The non-convex optimization of a PSGC will have many local minima due to the highly nonlinear response surface in a high dimensional parameter space. Stochastic methods, such as basin hopping, differential evolution, and stimulated annealing, have been developed to find approximate global optima instead of precise local optima. However, the stochastic algorithms require random sampling of the parameter spaces, and the performance is improved as a larger portion of the parameter space can be sampled. For the PSGC optimization, the cost of random sampling is prohibitive because 3D FDTD simulations are very expensive, and the high dimensional parameter space of PSGCs cannot be efficiently sampled due to curse of dimensionality. In this study, we tackle the non-convex optimization problem with a two-step deterministic approach: first to find a physics-based PSGC design within a reduced parameter space of a few most critical design parameters; second to solve the local optimization problem in the full parameter space with the physics-based PSGC as initial condition.

In the first step, we try to find a physics-based initial PSGC design by sweeping a few critical design parameters, including the grating pitch, fill factor, lattice tilt angle, the two linear terms $C_{00}$ and $C_{10}$ of the Chebyshev polynomials for the AspectRatio profile, and the 4$^{th}$ order cosine coefficient for the polygon shape. This process can be regarded as "forward design" since no gradient information is used, in contrast to the gradient-based "inverse design" by adjoint method. The grating pitch is dictated by phase matching conditions of the grating coupler as in eqn. (5), where $\lambda$ is the wavelength, $n_{eff}$ the grating effective index, $n_f$ the fiber mode index, and $\theta$ the beam angle. Assuming $n_{eff}$ of 2.85 and fiber mode index $n_f$ of 1.46, the calculated initial pitch value is about 495 nm.

$$pitch = \frac{\lambda}{n_{eff} - n_f \sin\theta} \tag{5}$$

Fill Factor determines both phase matching conditions and scattering strength. A rule of thumb is that the scattering strength is maximized at Fill Factor ~50%, which we will choose as the initial value for the parameter sweep. The lattice tilt angle $\gamma$ is also dictated by phase matching condition as in eqn. (6), and the calculated lattice tilt angle is about 3.11°.

$$\gamma = \sin^{-1}(\frac{pitch}{\lambda} n_f \sin\theta \sin\frac{\pi}{4}) \tag{6}$$

The two linear terms $C_{00}$ and $C_{10}$ of the Chebyshev polynomials for the AspectRatio determine the apodization profile. A fair starting point is to assume no apodization at all, or $C_{00}$=1.0 and $C_{10}$=0.0. The 4$^{th}$ order cosine coefficient for the polygon shape determines both PDW/PDL and the scattering strength, and without further quantitative information we will sweep the parameter around $c_4$=0 for all polygons across the grating area. We sweep each of the critical parameters around the initial guess values to obtain a physics-based PSGC design, and the optimal values of the parametric sweeps are pitch=501.1 nm, Fill Factor=0.5312, lattice tilt angle=2.666°, $C_{00}$=1.05, and $C_{10}$=-0.25 and the 4$^{th}$ order cosine coefficient for the polygon shape $c_4$=40 nm. This physics-based initial PSGC design is optimized in the parameter space of the few critical design parameters.

In the second step, we use the physics-based PSGC design obtained in the first step as the initial condition to solve local optimization problem in the full parameter space. We did not vary the initial PSGC design, which is optimized fairly well within the reduced space of the few critical design parameters. Instead, we control the optimizer setting to explore different convergence paths that all start with the same physics-based PSGC design. EMopt uses SciPy's optimize module as the backend optimizer, and we choose the quasi-Newton methods to find local minima of the FOM function: the exact Jacobian (gradient) is calculated by the adjoint-method with a forward FDTD and a reverse FDTD; an approximate Hessian is estimated from the Jacobians and design parameters of the current and previous iterations; an update step to the current design is calculated by solving a second-order approximation of the FOM function. In determining the update step, we choose the trust-region method instead of the line-search method in SciPy's optimize module, for its better handling of singular and non-positive-definite Hessians, faster convergence rate and greater flexibility to control the convergence behavior [17]. Damped BFGS method is selected to update the Hessian of the FOM with respect to the design parameters. Several adjoint optimizations with a wide range of initial trust-region radius are run in parallel to explore different convergence paths, and phase portraits of the Hessian eigenvalues and update steps are used to monitor the convergence behavior of the quasi-Newton method.  Convergence studies were performed to strike a subtle balance between accuracy and speed.  Figure 4 shows EMopt's FDTD simulation results of the same PSGC geometry at various mesh grid sizes of 30, 25, 20, and 15 nm.  Note that EMopt's FDTD solver uses monochromatic sources to solve Maxwell's equations only at the 3 individual wavelength points at 1305/1310/1315 nm, and the 3 calculated points on each plot are connected by quadratic fitting curves to show the trend of the spectra, which are not the actual spectra from EMopt's FDTD simulations.  As the FDTD mesh grid decreases from 30 nm to 15 nm, the peak loss deteriorates by less than a tenth of a dB, but the peak wavelength blueshifts by ~5 nm.  This shift in peak wavelength is due to the different meshing of the same geometry, which results in a change in effective etch depth.  This phenomenon is common in grating simulations and can be easily corrected by increasing the grating pitch.  The last subplot of Fig. 4 shows that after increasing the grating pitch parameter by 2 nm to 503.1 nm while leaving all other design parameters unchanged, the peak wavelength is brought back to 1310 nm, without significant loss penalty.

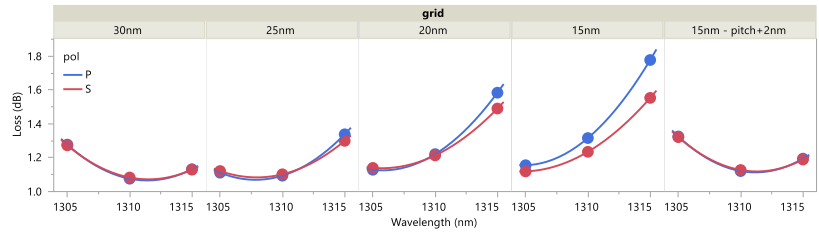

Fig. 4 EMopt simulated grating losses of the same PSGC at various grid sizes at 1305/1310/1315 nm, which are overlaid with quadratic fitting curves.

The grid size was set to 30 nm for the forward and reverse FDTD simulations performed by EMopt's Maxwell solver.  To achieve differentiable mapping from continuous material boundaries to discrete mesh grids, EMopt calculates effective permittivity of a grid cell as the volume-weighted average permittivity of all the materials that intersect the grid cell [18].  The adjoint-method calculated gradient vector was checked against the finite-difference method calculated gradient vector using the built-in tool of EMopt, and the relative error was less than 3% in the worst case.

The adjoint optimization was run on an HPE Superdome Flex S280 server with 16× 18-core Intel Xeon CPUs.  A total of 201 iterations were finished in about 19 days.  After 201 iterations, the coupling efficiency at 1310 nm reaches 77% (~1.14 dB) with worst case PDL below 0.01 dB across the 3 simulated wavelength points.  Figure 5 shows the S- and P-polarization coupling efficiencies at the three wavelength points of 1305nm, 1310nm, and 1315 nm, respectively.

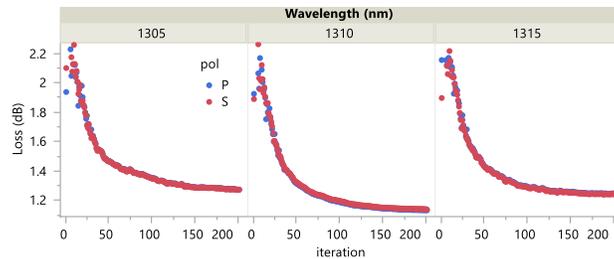

Fig. 5 Adjoint-optimized PSGC loss vs. iteration at the 3 wavelength points.

All 325 design parameters at all the 201 iterations were exported during the adjoint optimization to reconstruct and validate the adjoint-optimized PSGC designs in Lumerical FDTD Solutions.  Figure 6(a) shows the difference in peak losses simulated by EMopt and Lumerical. EMopt-Lumerical discrepancy is as little as 0.02/0.01 dB at iteration 1 for S- and

P-pol respectively but becomes more significant and reaches 0.08/0.07 dB at iteration 201, which is attributed to different meshing and index averaging on material interfaces between EMopt and Lumerical. Figure 6(b) shows the Lumerical-simulated coupling loss spectra of the final PSGC design. Peak wavelength shifts to 1305 nm in Lumerical due to the artifact of different meshing, which results in different effective etch depth. The 1dB bandwidth is ~27 nm. PDL of the final design, which is defined as the difference between S- and P-pol coupling losses, is plotted in Fig. 6(c). Within the 1 dB band from 1292 nm to 1319 nm, the worst case PDL is ~0.3 dB. To put these numbers in perspective, a comprehensive comparison of PSGC performances, including peak loss and PDL, can be found in [4]. The peak loss number achieved in this work is, to the best of our knowledge, the best performance of PSGCs without a bottom reflector.

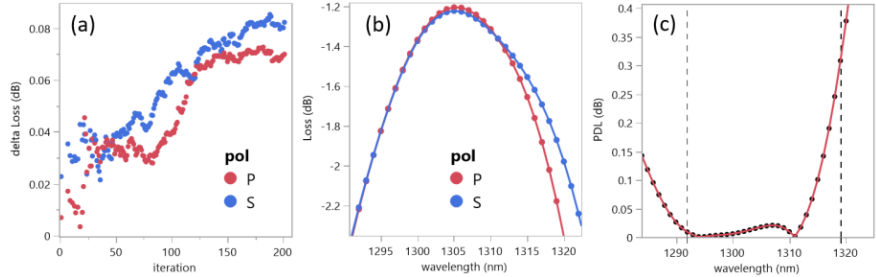

Fig. 6 (a) Difference in EMopt-Lumerical simulated PSGC loss. (b) Lumerical-simulated PSGC loss spectra of iteration 201. (c) Lumerical-simulated PSGC PDL spectra of iteration 201.

We strive to understand the rationales of the adjoint optimized PSGC, such that all the knowledge gained in this work is not lost when porting the design to other dielectric platforms or technology nodes. Figure 7(a) shows the top-view layout of the adjoint-optimized PSGC, and Figure 7(b) shows the contour map of the aspect ratio across the grating area. The adjoint-optimized PSGC design is consistent with our intuition in that the aspect ratio is higher at the edge of the grating for lower scattering strength and decreases toward the center of the grating for higher scattering strength. As light crosses the diagonal axis, the aspect ratio increases again to stretch the scattering elements along the orthogonal waveguide such that the scattering strength continues increasing, and as much power as possible can be extracted before light passes through the grating.

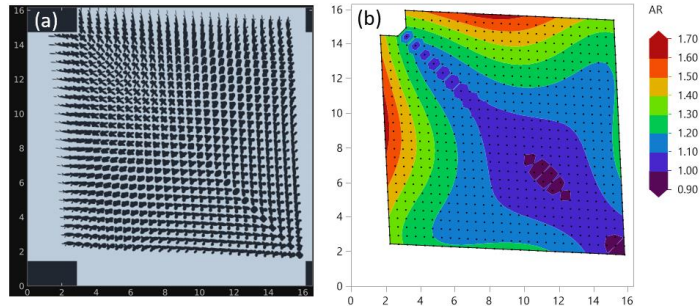

Fig. 7 (a) Top-view layout of the adjoint-optimized PSGC. (b) Contour map of the aspect ratio

Figure 8 shows the contour maps of the 6 sine and 6 cosine coefficients of the adjoint-optimized PSGC design, in micrometer and transformed by $10\times\log_{10}(|\bullet|)$. Note that on the color scale, -10 dB and -20 dB correspond to 100 nm and 10 nm, respectively. Given pitch=501.1 nm and fill factor=53.12%, each scattering element has the same area as that of a circle with radius=205 nm, so smaller trigonometric coefficients than 10 nm do not contribute much to the shape of the synthesized scattering element. The 5$^{th}$ and 6$^{th}$ order sine and cosine coefficients are mostly below 10 nm at the grating center, where the field intensity is higher and contributes more to the mode overlap. In the following we will disregard the 5$^{th}$ and 6$^{th}$ order sine and cosine functions.

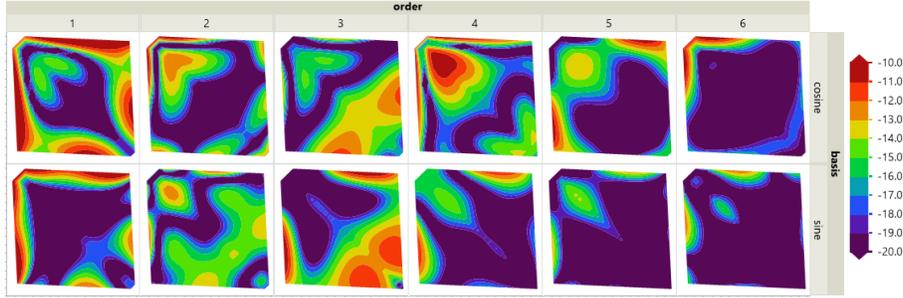

Fig. 8 Contour maps of the 12 trigonometric coefficients on log scale.

The higher amplitudes of the trigonometric coefficients on the starting edges, especially the 1$^{st}$ order sine and cosine, together with the higher aspect ratio, attempt to severely distort the scattering elements to form dendrites. These dendritic cells on the edges have weak scattering strength and serve as mode transition zone between the slab and the grating. At the grating center, the 2$^{nd}$ order sine/cosine, the 3$^{rd}$ order sine/cosine, and the 4$^{th}$ order cosine coefficients have strip-like features that are perpendicular to the diagonal axis of mirror symmetry, which inspire us to divide the center area of the grating into 3 zones from northwest to southeast, as shown in Figure 9. Within each of the 3 zones, a square area of 9 scatterers is zoomed in and shown in the insets.

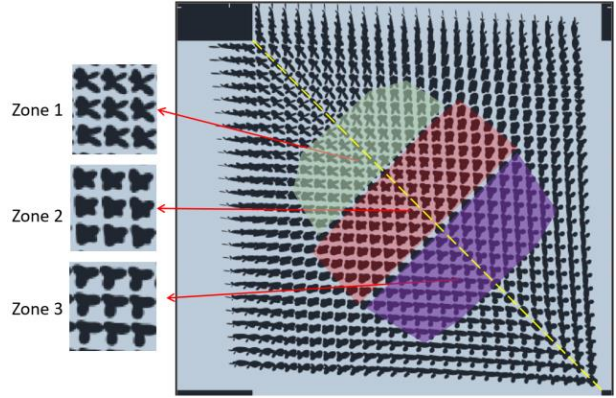

Fig. 9 Zoning of the grating center, with zoomed-in scatterer shapes in each zone.

We examine three simplified scatterers from the three zones respectively, to exemplify the impact of scatterer shape on the optical performance. The simplified scatterers are drawn by 4$^{th}$ order cosine and 2$^{nd}$ order sine in Zone 1 with $c_4$=0.1 um and $s_2$=0.045 um, by 4$^{th}$ order cosine only in Zone 2 with $c_4$=0.04 um, and by 3$^{rd}$ order cosine and 2$^{nd}$/3$^{rd}$ order sine in Zone 3 with $c_3$=-0.045 um, $s_2$=-0.05 um, and $s_3$=-0.05 um. The simplified scattering elements are shown in Figure 10(a). To study the characteristics of each of the 3 exemplar scatterers, we simulate 3 uniform PSGCs that have homogeneous scattering elements within the entire grating area for the 3 exemplar scattering elements, respectively. Figure 10(b) shows coupling efficiency spectra for the 3 PSGC designs. The scattering elements in zones 1-3 have negative, positive, and close to zero PDW respectively, which is defined as S-pol peak wavelength minus P-pol peak wavelength.

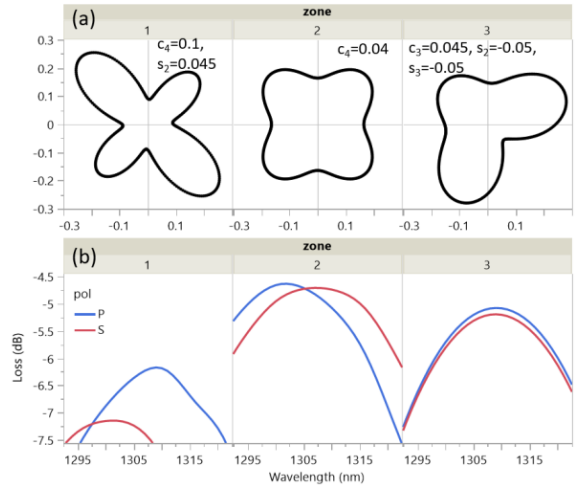

Fig. 10 (a) Exemplar scatterers from zones 1 to 3, respectively. (b) Loss spectra of PSGCs comprising homogeneous scattering elements of the 3 types.

Figure 11 shows the amplitude of the Ey field at 1310 nm, along the X-axis at the center of the horizontal waveguide, for the 3 exemplary PSGCs that comprise homogeneous scattering elements of the 3 zones. The scattering element in zone 1 has the weakest scattering while the scattering element in zone 3 has the strongest scattering.

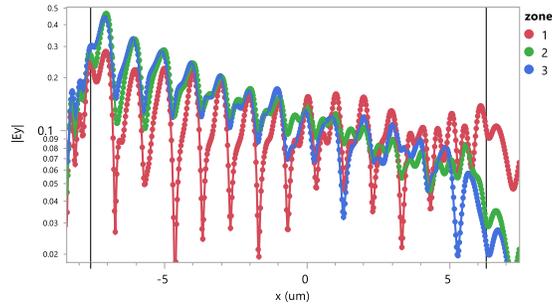

Fig. 11 Ey field at 1310nm at the center of the horizontal waveguide

With above information, we can extract the concept of a 3-zone PSGC design as shown schematically in Figure 12. One counter-intuitive observation is that the PDW does not have to be small for all regions in the grating. A designer can freely choose the scatterer shape to have combinations of positive- and negative-PDW regions and still get zero overall PDW. Despite the infinite number of combinations to achieve zero overall PDW, an obvious constraint is the requirement for optimal scattering strength, which is weaker at the beginning edges and stronger at the center. The adjoint-optimized PSGC in this work may not be the global optimum, and better solutions of more sophisticated designs can be achieved, for example, by tuning the PDL weight $\lambda$ in the FOM, or by increasing the order of basis functions. Instead, the results of adjoint optimization proved that our proposed framework is highly effective in designing scatterer shape to pursue the dual mandate of low loss and low PDL simultaneously.

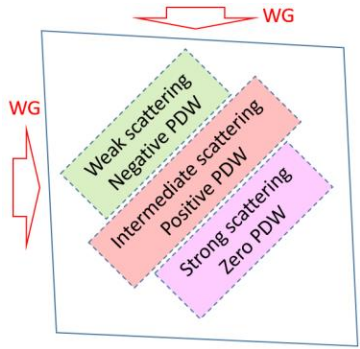

Fig. 12. Concept of the 3-zone PSGC design.

## 3. Design considerations

Despite the success of the adjoint method in exploring high dimensional parameter spaces that are otherwise intractable, tested optical performance of photonic devices on silicon will inevitably shift from numerical simulations, due to limitations of electromagnetic simulations, and biases in microfabrication processes. It is therefore important to experimentally validate the adjoint-optimized designs by taping out and testing DOEs that consist of variations of design parameters to account for the discrepancies between simulations and test data. As the number of design parameters increase, it becomes increasingly difficult to study the interaction of the design parameters with limited resources of die area and testing time [12][19].

The large number of parameters in the PSGC design is due to a combination of high order of trigonometric basis functions crossed with high order of 2D Chebyshev polynomials, so we examine the choice of the orders of bases. As shown in Fig. 8 in Section 2, the $5^{th}$ and $6^{th}$ order sine and cosine functions do not contribute significantly to the shape of the scatterers and will be disregarded. As the order of trigonometric coefficients decreases from 6 to 4 while the order of Chebyshev polynomials remains at 5, the adjoint-optimized PSGC loss is higher by ~0.13 dB. We reran adjoint optimizations with the first 4 orders of sine and cosine functions but reduced the order of Chebyshev polynomials from 5 to 2 in each optimization. Figure 13 shows the contour maps of the trigonometric coefficients with various order of Chebyshev polynomials. Despite the disappearance of many fine structures in the contour maps as the order decreases, the strip-like zone structures at the grating center persist in the contour maps of trigonometric coefficients. For example, the $4^{th}$ order cosine coefficient always decreases from zone 1 to zone 3, and the $2^{nd}$ order sine coefficient decreases and changes sign from zone 1 to zone 3. We surmise that the same strip-like structures will persist if the optimization problem changes to other orthonormal polynomial basis functions, such as Hermite or Legendre polynomials. Figure 14 shows the optimized peak loss at 1310nm, plotted versus the order of the 2D Chebyshev polynomials. The coupling loss increases by ~0.3 dB as the order of Chebyshev polynomials decreases from 5 to 2.

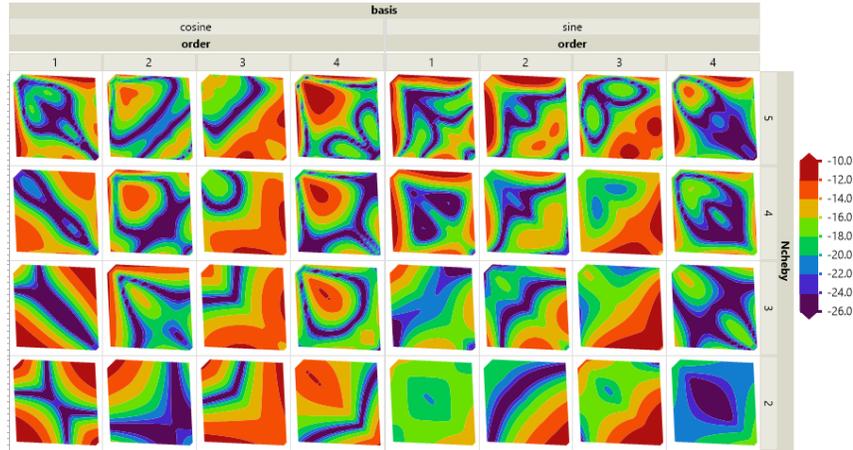

Fig. 13 Contour maps of trigonometric coefficients with Chebyshev polynomial order from 5 to 2

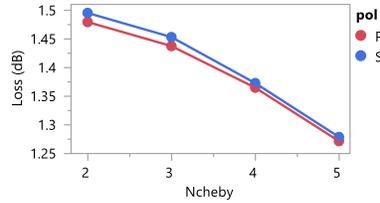

Fig. 14 Peak loss vs. the order of Chebyshev polynomials

Another practical issue of the adjoint-optimized photonic devices is the manufacturability. In this study, while utilizing shape-based optimization, we chose unconstrained, parametric geometry-based optimization to explore different PSGC topologies, which is a key benefit that is normally of topology-based optimization methods. For example, the scatterers at the edge of the PSGC are stretched to merge with neighboring cells to achieve lower/higher scattering strength or to reduce reflection at the slab-grating interface. The unconstrained adjoint-optimization of PSGCs results in many features that are beyond manufacturability limits. We study the impact of one of the most prevalent non-manufacturable features, namely the fine Si lines that are below a certain threshold. The adjoint optimization is modified to detect minimal spacing between neighboring cells that is narrower than a threshold, which will unlikely survive the lithography and/or subsequent etching processes. To imitate the loss of such fine Si lines, the geometry generation script automatically patches all the fine Si lines that are narrower than the threshold with $SiO_2$ rectangles, as shown in the Fig. 15. Note that the operation of trimming fine Si lines will distort the response surface of PSGC optimization and create local minima, where differentiability is no longer preserved.

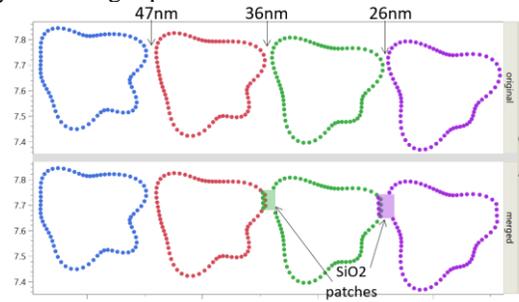

Fig. 15 Scatterers of unconstrained optimization (top row) and the optimization that imitates the loss of fine features (bottom row)

Adjoint optimization of the original PSGC design (6 sine + 6 cosine basis functions × 5[th] order Chebyshev polynomials) is launched with minimum Si line width of 5 nm, 25 nm, and 45 nm, respectively. The 1310 nm S- and P-pol losses of the original optimization, as well as the modified optimizations of the three minimal Si line widths are shown in Figure 16. As the minimal Si line width increases from 0 to 45 nm, the adjoint-optimized PSGC suffers a loss penalty of ~0.1 dB. It is also worth noting that as the minimal Si line width increases and more Si is trimmed, the optimization terminates earlier as the response surfaces are artificially distorted with more local minima, since the lost Si features will no longer respond to perturbations. Instead of the hard constraint of minimum feature size that sacrifices differentiability of the FOM function, numerous prior works have also demonstrated adjoint-optimizations with differentiable penalty functions [20][21][22], which will be an interesting topic for further study.

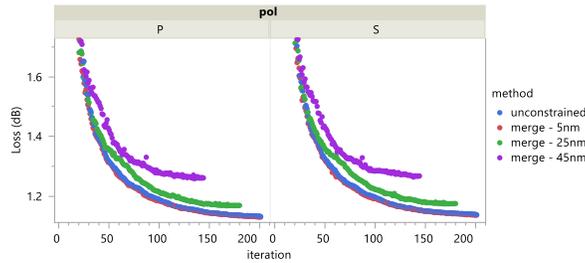

Fig. 16 Adjoint-optimized PSGC peak losses of the unconstrained design, and the design with minimum Si line width of 5 nm, 25 nm, and 45 nm

Besides the loss of fine Si features that is discussed above, we also simulated the impact of two types of fabrication variation, namely the etch depth and the lithography exposure dose. To imitate over- and under-exposure of photolithography, we vary the PSGC's Fill Factor parameter by +/-0.03 around the nominal value of 0.5312. We also vary the etch depth by +/- 5 nm around the nominal etch depth of 140 nm. The fabrication variations in exposure dose and etch depth are applied to two PSGC designs: a simple baseline PSGC with the 4[th] order cosine coefficient $c_4$=40 nm and AspectRatio=1.0 across the entire grating area, and the complicated adjoint-optimized PSGC. Simulation results show that the adjoint-optimized PSGC has slightly higher sensitivity to fabrication errors as the simple baseline PSGC at the process corners: per 0.03 change in the Fill Factor, the peak loss may change by 0.1 dB and peak wavelength by 5 nm; per 5 nm change in the etch depth, the peak loss may change by 0.1dB and peak wavelength by 8 nm. Prior studies [12][21] also confirmed that adjoint-optimized devices with complicated geometries can have similar sensitivities to process variations as simple designs by conventional methods. Combining the impact of etch depth and exposure dose variations, we expect the adjoint-optimized PSGC to have wafer-level loss spread of 0.3~0.4 dB, and wavelength spread of 15~20 nm.

## 4. Discussion

Chebyshev polynomials are general orthonormal basis functions that can approximate arbitrary continuous functions. They were chosen to synthesize the trigonometric coefficients and the aspect ratio when we had limited knowledge of the desired target functions before we started the adjoint optimization. In hindsight, Chebyshev polynomials might not be the optimal choice of basis functions for the PSGC problem, as more geometrical and topological structures are unveiled in the adjoint-optimized PSGC designs. We explored the change of basis with Singular Value Decomposition (SVD), which transform the coefficient matrices of 2D Chebyshev polynomials with two unitary matrices $U$ and $V^h$ to result in one diagonal matrix of singular value $S$. The SVD transformation essentially chooses a more relevant set of basis functions, which are linear combinations of the original Chebyshev polynomial bases. As expected, the SVD transformed basis functions also have similar strip-like structures at the grating center. Other techniques for dimensionality reduction, such as PCA, can also be incorporated in the design of complex devices [23][24].

We conducted a top-down analysis by breaking down the PSGC loss. In Lumerical validation of the adjoint-optimized PSGC, we enclose the grating area with a box of 6 monitors, and the PSGC loss is decomposed into the sum of 4 components: fiber-grating mode overlap, reflection to the waveguide, leakage to the Si substrate, and leakage that passes through the grating. Figure 17 shows how the 4 loss components evolved as the adjoint optimization progressed, where iteration 1 is the initial PSGC designed by simple parameter sweeping, and iteration 201 is the adjoint-optimized PSGC. Mode overlap is improved the most by a total of 0.45dB; reflection is improved by 0.15 dB, likely by the mode matching zone that consists of the dendritic cells between the unetched slab and the grating; leakage passing through the grating is slightly improved by 0.05 dB, likely by higher order terms in the aspect ratio profile; leakage to the substrate is not improved by the grating design and stands out as the most significant component in the final adjoint-optimized design. Efforts to further improve mode overlap (for example with sub-wavelength grating or multi-etching [25]), or to reduce through leakage will likely have diminishing returns. The most effective method to further improve the adjoint-optimized PSGC is to use a bottom reflector to reduce leakage to the Si substrate [26-29].

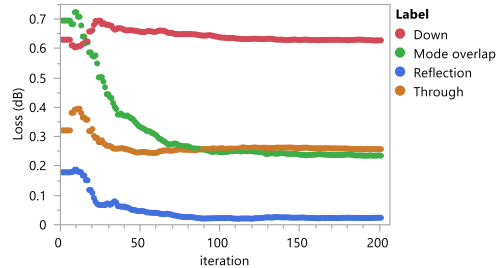

Fig. 17 Loss breakdown of the adjoint-optimized PSGCs.

In this work, we chose shape-based optimization over topology-based optimization, such that valuable physical insights can be extracted from the adjoint optimization and used to port the PSGC design to other technology or dielectric platforms with a minimum of effort. We formulated the problem as shape-based optimization but let the scatterer shape to freely change without any constraints. The unconstrained shape-based optimization achieved similar effect as topology-based optimization, which is evident from the large number of connected scatterers in the PSGC layout. If we examine the spacing between scatterers of the adjoint-optimized PSGC, only 360 of the total 775 scatterers at the grating center are standalone and not connected to neighboring scatterers. The maximum spacing between standalone scatterers is 77nm, and the maximum area overlap ratio between merged scatterers is ~60%. These numbers imply that modulating the aspect ratio of individual scatterers may not be the optimal choice to achieve the desired scattering strength under the constraints of manufacturability. We might have to consider other architecture to better respect the design rules. Also, the high area overlap ratio between merged scatterers, especially those on the edges, imply that the fill factor is not constant and phase matching conditions are no longer preserved. The fill factor needs to be included in the adjoint optimization to restore phase matching for the connected edge cells, but the gain in the coupling efficiency will not be more than a few tenths of a dB.

**Funding**

Hewlett Packard Enterprise.

## Disclosures

The authors declare that there are no conflicts of interest related to this article.

## Data availability

Data underlying the results presented in this paper are not publicly available at this time but may be obtained from the authors upon reasonable request.

## References

1. D. Taillaert, H. Chong, P. I. Borel, L.H. Frandsen, R.M. De La Rue, and R. Baets, "A compact two-dimensional grating coupler used as a polarization splitter," IEEE Photon. Technol. Lett. **15**, pp. 1249-1251 (2003).
2. F. Van Laere, W. Bogaerts, P. Dumon, G. Roelkens, D. Van Thourhout, and R. Baets, "Focusing polarization diversity gratings for silicon-on-insulator integrated circuits," 2008 5th IEEE Intl. Conf. Group IV Photon., 2008, pp. 203-205 (2008).
3. L. Carroll, D. Gerace, I. Cristiani, S. Menezo, and L.C. Andreani, "Broad parameter optimization of polarization diversity 2D grating couplers for silicon photonicsics," Opt. Express. **21**(18), pp. 21556-21568 (2013).
4. L. Cheng, S. Mao, Z. Li, Y. Han, and H.Y. Fu, "Grating couplers on silicon photonics: design principles, emerging trends and practical issues," Micromachines 11(7), pp. 666, (2020).
5. A. Mekis, S. Gloeckner, G. Masini, A. Narasimha, T. Pinguet, S. Sahni, and P. De Dobbelaere, "A grating-coupler-enabled CMOS photonics platform," IEEE J. Sel. Topics Quantum Electron., 17(3), pp.597-608 (2011).
6. L. Verslegers, A. Mekis, T. Pinguet, Y. Chi, G. Masini, P. Sun, A. Ayazi, K.Y. Hon, S. Sahni, S. Gloeckner, C. Baudot, F. Boeuf, and P. De Dobbelaere, "Design of low-loss polarization splitting grating couplers," Adv. Photon. Commun., OSA Technical Digest, paper JT4A.2, (2014).
7. S. Plantier, D. Fowler, K. Hassan, O. Lemonnier, R. Orobtchouk, "Impact of scattering element shape on polarization dependent loss in two dimensional grating couplers," 2016 13th IEEE Intl. Conf. Group IV Photon., pp. 76-77, (2016).
8. Y. Sobu, S.-H. Jeong, and Y. Tanaka, "Demonstration of low polarization dependent loss of 1.3 um two dimensional grating coupler," 2017 IEEE 14th Int. Conf. Group IV Photon., pp.127-128, (2017).
9. Y. Xue, H. Chen, Y. Bao, J. Dong, and X. Zhang, "Two-dimensional silicon photonic grating coupler with low polarization-dependent loss and high tolerance," Opt. Express 27(16), pp.22268-22274 (2019).
10. C. M. Lalau-Keraly, S. Bhargava, O. D. Miller, and E. Yablonovitch, "Adjoint shape optimization applied to electromagnetic design," Opt. Express 21(18), pp. 21693-21701 (2013).
11. A. Michaels and E. Yablonovitch, "Inverse design of near unity efficiency perfectly vertical grating couplers," Opt. Express 26(4), pp.4766-4779 (2018).
12. P. Sun, T. Van Vaerenbergh, M. Fiorentino, and R. Beausoleil, "Adjoint-method-inspired grating couplers for CWDM O-band applications," Opt. Express 28(3), pp.3756-3767 (2020).
13. S. Hooten, T. Van Vaerenbergh, P. Sun, S. Mathai, Z. Huang, and R. G. Beausoleil, "Adjoint optimization of efficient CMOS-compatible Si-SiN vertical grating couplers for DWDM applications," IEEE J. Lightwave Technol. 38(13), pp.3422-3430, (2020).
14. I.P. Natanson, *Constructive Function Theory*, (Ungar, 1964), Vol. 1, pp. 53.
15. A. Michaels, "EMopt," May 2019, Electromagnetic optimization software available at https://github.com/anstmichaels/emopt.
16. S. Boyd, and L. Vandenberghe, *Convex Optimization*, (Cambridge University Press, 2004), pp. 72.
17. J. Nocedal, and S. Wright, *Numerical Optimization*, (Springer, 2006), 2nd edition, Chap. 6.2.
18. A. Michaels, and E. Yablonovitch, "Leveraging continuous material averaging for inverse electromagnetic design," Opt. Express **26**(24), pp. 31717-31737, (2018).
19. T. Van Vaerenbergh, P. Sun. S. Hooten, M. Jain, Q. Wilmart, A. Seyedi, Z. Huang, M. Fiorentino, and R. Beausoleil, "Wafer-level testing of inverse-designed and adjoint-inspired vertical grating coupler designs compatible with DUV lithography," Opt. Express 29(23), pp.37021-37036 (2021).
20. A. Michaels, M. C. Wu, and E. Yablonovitch, "Hierarchical design and optimization of silicon photonics," IEEE J. Sel. Topics Quantum Electron. 26(2), pp. 8200512, (2020).
21. A.M. Hammond, J.B. Slaby, M.J. Probst, and S.E. Ralph, "Multi-layer inverse design of vertical grating couplers for high-density, commercial foundry interconnects," Opt. Express 30(17), pp. 31058-31072 (2022).
22. T. Van Vaerenbergh, S. Hooten, M. Jain, P. Sun, Q. Wilmart, A. Seyedi, Z. Huang, M. Fiorentino, and R. Beausoleil, "Wafer-level testing of inverse-designed and adjoint-inspired dual layer Si-SiN vertical grating couplers," J. Phys. Photonics 4, 044001, (2022).
23. D. Melati, Y. Grinberg, M. K. Dezfouli, S. Janz, P. Cheben, J. H. Schmid, A. Sanchez-Postigo, and D.-X. Xu, "Mapping the global design space of nanophotonic components using machine learning pattern recognition," Nature Commun. 10, 4775, (2019).

24. M. Al-Digeil, Y. Grinberg, D. Melati, M. K. Dezfouli, J. H. Schmid, P. Cheben, S. Janz, and D.-X. Xu, "PCA-boosted autoencoders for nonlinear dimensionality reduction in low data regimes," arXiv:2205.11673
25. T. Watanabe, Y. Fedoryshyn, J. Leuthold, "2-D grating couplers for vertical fiber coupling in two polarizations," IEEE Photon. J. 11(4), pp.1-9, (2019).
26. L. Carroll, D. Gerace, I. Cristiani, and L.C. Andreani, "Optimizing polarization-diversity couplers for Si-photonics: reaching the -1dB coupling efficiency threshold," Opt. Express 22(12), pp.14769-14781 (2014).
27. M.F. Rosa, P. de la Torre Castro, N. Hoppe, L. Rathgeber, W. Vogel, and Manfred Berroth, "Novel design of two-dimensional grating couplers with backside metal mirror in 250 nm silicon-on-insulator," 2017 Int. Conf. Num. Sim. Optoelectron. Dev., pp.81-82, (2017).
28. Y. Luo, Z. Nong, S. Gao, H. Huang, Y. Zhu, L. Liu, L. Zhou, J. Xu, L. Liu, S. Yu, and X. Cai, "Low-loss two-dimensional silicon photonic grating coupler with a backside metal mirror," Opt. Lett. 43(3), pp.474-477, (2018).
29. B. Chen, X. Zhang, X. Wen, Z. Ruan, Y. Zhu, L. Liu, "High efficiency and polarization insensitive two-dimensional grating coupler on silicon," 2018 Asia Commun. Photon. Conf., pp.1-2, (2018).